\documentclass[twocolumn,prd,nofootinbib]{revtex4}
\usepackage{amsmath}
\usepackage{graphicx}

\begin{document}

\title{Brane-world cosmology with black strings }
\author{L\'{a}szl\'{o} \'{A}. Gergely}
\affiliation{Departments of Theoretical and Experimental Physics, University of Szeged,
Hungary}

\begin{abstract}
We consider the simplest scenario when black strings / cigars penetrate the
cosmological brane. As a result, the brane has a Swiss-cheese structure,
with Schwarzschild black holes immersed in a Friedmann-Lema\^{\i}%
tre-Robertson-Walker brane. There is no dark radiation in the model, the
cosmological regions of the brane are characterized by a cosmological
constant $\Lambda $ and flat spatial sections. Regardless of the value of $%
\Lambda $, these brane-world universes forever expand and forever
decelerate. The totality of source terms in the modified Einstein equation
sum up to a dust, establishing a formal equivalence with the general
relativistic Einstein-Straus model. However in this brane-world scenario
with black strings the evolution of the cosmological fluid strongly depends
on $\Lambda $. For $\Lambda \leq 0$ it has positive energy density $\rho $
and negative pressure $p$ and at late times it behaves as in the
Einstein-Straus model.\ For (not too high ) positive values of $\Lambda $
the cosmological evolution begins with positive $\rho $ and negative $p$,
but this is followed by an epoch with \textit{both} $\rho $ and $p$
positive. Eventually, $\rho $ becomes negative, while $p$ stays positive. A
similar evolution is present for high positive values of $\Lambda $, however
in this case the evolution ends in a \textit{pressure singularity},
accompanied by a \textit{regular} behaviour of the cosmic acceleration. This
is a novel type of singularity appearing in brane-worlds.
\end{abstract}

\date{\today }
\startpage{1}
\endpage{}
\maketitle

\section{Introduction}

There is increasing expectation that general relativity is only the
low-energy effective theory of gravitation. This belief originates mainly in
the lack of the singularity avoidance of the theory and its still existing
failure to be merged with quantum principles. String theory suggests that
the five-dimensional Einstein gravity containing our four-dimensional (4D)
world as a brane \cite{ADD} - \cite{RS2} may be a better approximation. In
the original Randall-Sundrum (RS) setup \cite{RS2}, a flat brane was
symmetrically embedded into a 5-dimensional anti de Sitter (AdS5) space-time
(the bulk). Later on, generalizations of the model to curved branes and more
generic bulk space-times were advanced. In such models, on the brane,
gravity is governed by a modified Einstein equation \cite{SMS}: 
\begin{equation}
G_{ab}=-\Lambda g_{ab}+\kappa ^{2}T_{ab}+\widetilde{\kappa }^{4}S_{ab}-%
\mathcal{E}_{ab}\ ,  \label{modE}
\end{equation}%
where $S_{ab}$ denotes a quadratic expression in the energy-momentum tensor $%
T_{ab}$: 
\begin{equation}
S_{ab}={{\frac{1}{12}}}TT_{ab}-{{\frac{1}{4}}}T_{ac}T^{c}{}_{b}+{{\frac{1}{24%
}}}g_{ab}\left( 3T_{cd}T^{cd}-T^{2}\right) \,,  \label{S}
\end{equation}%
and $\mathcal{E}_{ab}$ represents the electric part of the Weyl curvature of
the bulk. The brane gravitational constant $\kappa ^{2}$ and the brane
cosmological constant\ $\Lambda $ are related to the bulk gravitational
constant $\widetilde{\kappa }^{2}$, bulk cosmological constant $\widetilde{%
\Lambda }$ and the (positive) brane tension $\lambda $ through 
\begin{eqnarray}
6\kappa ^{2} &=&\widetilde{\kappa }^{4}\lambda \ , \\
2\Lambda &=&\kappa ^{2}\lambda +\widetilde{\kappa }^{2}\widetilde{\Lambda }\
.
\end{eqnarray}%
In such a scenario our observable universe can be imagined as a
Friedmann-Lema\^{\i}tre-Robertson-Walker (FLRW) brane moving in a static
Schwarzschild-anti de Sitter (SAdS5) background. Cosmology is modified
accordingly, however there is no room left for brane-black holes in this
model, which are merely test particles on the brane.

The most simple extension of black holes in higher-dimensional gravity is
represented by black strings. If black strings populate the bulk, then they
penetrate the brane and Schwarzschild black holes emerge. On the brane,
black holes could be introduced in the most simple way by a cut and paste
method. Replacing spheres removed from the FLRW brane with vacuum regions
centered around Schwarzschild black holes we obtain a Swiss-cheese model. In
principle such a model could be a more realistic description of the Universe
than the FLRW brane, as it contains Schwarzschild black holes with \textit{%
finite mass} in the cosmological background. The cosmological consequences
can be important, as for example the original general relativistic
Swiss-cheese model, the Einstein-Straus model \cite{ES} has predicted a
modified luminosity-redshift relation \cite{Kantowski}.

It is well known, that in the context of the generalized RS brane-worlds,
the generic \textit{spherically symmetric} black holes are characterized by
both a mass $m$ and a tidal charge $q$, the latter originating from the
electric part of the bulk Weyl curvature \cite{tidalRN}. Thus spherically
symmetric brane black holes resemble the Reissner-Nordstr\"{o}m solutions,
but the tidal charge can take negative values as well, in contrast with
general relativity, where $q=Q^{2}$ represents the square of the electric
charge $Q$. In spite of the fact that the bulk solution containing such a
brane is unknown, various astrophysical phenomena were already studied in
this scenario, like gravitational lensing \cite{MM}. The tidal charge term
represents a correction to the Schwarzschild potential, which scales as $%
r^{-2}$ ($r$ being the distance measured from the center of symmetry). This
correction term comes from the Weyl curvature of the bulk (from the $%
\mathcal{E}_{ab}$ source term) and it is in fact the most important piece of
information we have about the bulk containing the tidal charged brane black
hole.

We remark here that the perturbative analysis of the gravitational field of
a spherically symmetric source in the weak field limit, in the context of
the original RS setup has shown other type of corrections to the
Schwarzschild potential (which depend on the curvature radius of the fifth
dimension) \cite{RS2}, \cite{GT}-\cite{GKR}. These scale as $r^{-3}$, thus
they are different from the correction induced by the tidal charge in the
tidal charged black hole solution. This shows that the bulk containing the
tidal charged black hole brane, under no circumstances can\textit{\ }behave
as a linearly perturbed AdS5 space-time. This is the second information we
have about the bulk containing the tidal charged brane black hole.

The same holds true for the simplest black hole on the brane, given by the
Schwarzschild solution (with $q=0$). As the Schwarzschild potential does not
have any $r^{-3}$ correction term in this case, the bulk containing the
Schwarzschild brane black hole is again not a linearly perturbed AdS5
space-time. Remarkably, we do know more on the extension into the bulk of
the Schwarzschild brane black hole, than of the generic tidal charged black
hole. As conjectured in \cite{ChRH}, a Schwarzschild brane black hole can be
embedded in the bulk only by extending the singularity into the bulk.
Therefore a black string with singular AdS horizon emerges. Recently, the
gravity wave perturbations of such a black-string brane-world were computed 
\cite{SCMlet}. It is widely believed, that due to the Gregory-Laflamme
instability \cite{GL} the black string will shortly decay into a black cigar 
\cite{Gregory}. However, more recently it was shown that under very mild
assumptions, classical event horizons cannot pinch off \cite{HorowitzMaeda}.

\begin{figure}[tbp]
\includegraphics[height=6cm]{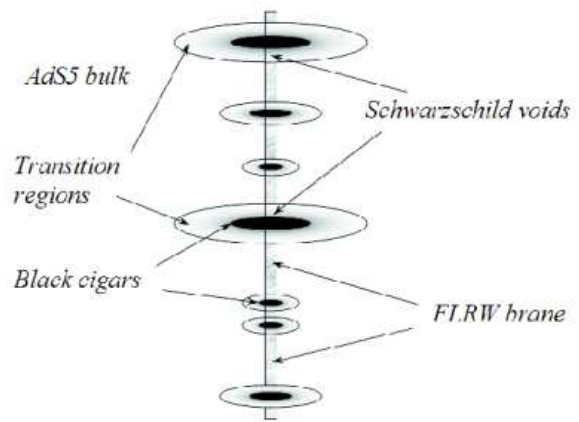}
\caption{(color online). A schematic, one-dimensional representation of a
Swiss-cheese brane-world. The FLRW brane is embedded in an AdS5 bulk. The
brane Schwarzschild black holes extend into the bulk as black strings /
cigars. A transition zone separates these black strings / cigars from the
AdS5 bulk regions.}
\label{Fig1}
\end{figure}

The configuration we propose to study in this paper is schematically
represented on Fig \ref{Fig1}. The brane is FLRW and it contains a number of
Schwarzschild voids. Irrespective of which region we are on the brane, the
electric part of the bulk Weyl curvature is switched off. While this is
customary for the cosmological regions (then the bulk becomes AdS5 and there
is no dark radiation on the respective brane regions), it does not represent
an obvious choice for black strings. As motivation, we note that stable
black string solutions with $\mathcal{E}_{ab}=0$ arise in the two brane
models, introduced in Ref. \cite{RS1}. Such black strings extending between
the physical brane and a "shadow brane" were already employed as a model of
brane-world black holes \cite{SCMlet}, and we follow the same choice here.

A difficult problem would be to find the global bulk solution for this
highly asymmetric configuration. We propose again the cut and paste method
here, by inserting a transition zone between the black string / cigar metric and the
AdS5 regions. As the exact shape of these regions is not fixed a priori, we
conjecture that by choosing an appropriate shape for these regions, the
boundary problem becomes solvable. We do not propose to further discuss this
topic here, but rather concentrate on the existence of the brane
cosmological solution, in the same spirit as brane stellar solutions were
discussed in \cite{ND}-\cite{HarkoMak}, weak gravitational lensing in \cite%
{MM}, strong lensing in \cite{Whisker} and galactic rotation curves in \cite%
{MakHarko}. 

In \cite{NoSwissCheese} the Swiss-cheese brane-world model was already
studied, assuming $\mathcal{E}_{ab}=0$. By allowing for a cosmological
constant $\Lambda $ in the FLRW regions the possibility of having different
brane tensions $\lambda $ in the FLRW and Schwarzschild regions of the
brane, and / or having different cosmological constants $\widetilde{\Lambda }
$ in the bulk regions surrounding them was raised. The junction conditions
on the brane have been derived and it has been shown that for \textit{static}
cosmological fluid no such brane (beside the empty one, the Minkowski brane)
can exist.

In Section 2 we present the simplest examples of \textit{dynamic}
Swiss-cheese universes on the brane. These consist of black strings / cigars
penetrating a cosmological FLRW brane without dark radiation. The
intersections appear on the brane as Schwarzschild spheres. We find that
contrarily to the static case, the modified Einstein equations admit brane
solutions of Swiss-cheese type. We study in detail the cosmological
evolution in these models in Section 3. Additional comments on these models,
including their classification with respect to the value of the cosmological
constant, and remarks on a novel type of pressure singularity emerging in
these Swiss-cheese models are presented in the Concluding Remarks.

\section{Swiss-cheese brane-worlds}

We construct the dynamic Swiss-cheese brane- by immersing 4D Schwarzschild
vacua with constant comoving radius $\chi _{0}$ in a \textit{non-static}
FLRW brane with spatially flat spatial sections ($k=0$), cosmological
constant $\Lambda $ and tension $\lambda $.

The cosmological evolution in the FLRW regions is governed by the
generalized Friedmann and the generalized Raychaudhuri equation \cite{Decomp}%
:%
\begin{gather}
\frac{\dot{a}^{2}}{a^{2}}=\frac{\Lambda }{3}+\frac{\kappa ^{2}\rho }{3}%
\left( 1+\frac{\rho }{2\lambda }\right) \ ,  \label{Fried} \\
\frac{\ddot{a}}{a}=\frac{\Lambda }{3}-\frac{\kappa ^{2}}{6}\left[ \rho
\left( 1+\frac{2\rho }{\lambda }\right) +3p\left( 1+\frac{\rho }{\lambda }%
\right) \right] \ .  \label{Raych}
\end{gather}%
General relativity is recovered for $\rho /\lambda \rightarrow 0$. The
continuity of the induced metric and of the extrinsic curvature on the
junction surfaces between the Schwarzschild vacua and the FLRW regions imply 
\cite{NoSwissCheese}%
\begin{eqnarray}
a\dot{a}^{2} &=&\frac{2m}{\chi _{0}^{3}}\ ,  \label{junct1} \\
a^{2}\ddot{a} &=&-\frac{m}{\chi _{0}^{3}}\ .  \label{junct2}
\end{eqnarray}%
Eq. (\ref{junct1}) can be easily integrated to obtain the evolution of the
scale factor $a$ in cosmological time $\tau $:%
\begin{equation}
a^{3}=\frac{9m\tau ^{2}}{2\chi _{0}^{3}}\ .  \label{atau}
\end{equation}%
Then the other junction condition is satisfied as well. These Swiss-cheese
universes therefore \textit{forever expand}:%
\begin{equation}
\frac{\dot{a}}{a}=\frac{2}{3\tau }\ ,  \label{adot}
\end{equation}%
\textit{forever decelerate:}%
\begin{equation}
\frac{\ddot{a}}{a}=-\frac{2}{9\tau ^{2}}\ ,  \label{adot2}
\end{equation}%
and their expansion ceases at $\tau \rightarrow \infty $, as expected for $%
k=0$. An integration constant was set to zero in Eq. (\ref{atau}) such that
the origin of the $\tau $ coordinate is at the Big Bang. From the junction
condition (\ref{junct1}) and the generalized Friedmann equation (\ref{Fried}%
), the mass of a Schwarzschild void with comoving radius $\chi _{0}$ emerges 
\cite{NoSwissCheese} in terms of the scale factor $a$ and density of the
cosmological fluid $\rho $ as:%
\begin{equation}
m=\frac{a^{3}\chi _{0}^{3}}{6}\left[ \Lambda +\kappa ^{2}\rho \left( 1+\frac{%
\rho }{2\lambda }\right) \right] \ .  \label{m}
\end{equation}

The junction condition (\ref{junct2}) and the generalized Raychaudhuri
equation (\ref{Raych}) give the equation of state of the fluid \cite%
{NoSwissCheese}. Taking into account Eq. (\ref{m}) this becomes%
\begin{equation}
a^{3}\left( \rho +p\right) \left( \rho +\lambda \right) =\frac{6m\lambda }{%
\kappa ^{2}\chi _{0}^{3}}\ .  \label{rhop2}
\end{equation}%
Comparison of Eqs. (\ref{atau}) and (\ref{m}) leads to%
\begin{equation}
\kappa ^{2}\rho \left( 1+\frac{\rho }{2\lambda }\right) =-\Lambda +\frac{4}{%
3\tau ^{2}}\ ,
\end{equation}%
with solutions 
\begin{equation}
\frac{\rho _{1,2}}{\lambda }=-1\pm \sqrt{1-\frac{2\Lambda }{\kappa
^{2}\lambda }+\frac{8}{3\kappa ^{2}\lambda \tau ^{2}}}\ .  \label{rhotau}
\end{equation}%
For a positive energy density we choose the $+$ solution and also 
\begin{equation}
3\Lambda \tau ^{2}<4
\end{equation}%
has to hold. This is immediate for $\Lambda \leq 0$. For any positive $%
\Lambda $ the energy density is positive for $\tau <\tau _{1}\equiv 2/\sqrt{%
3\Lambda }$ and negative afterwards. For a cosmological constant small
enough the positivity of $\rho $ is maintained for a long time even in this
case. However, for having $\rho $ real the condition 
\begin{equation}
3\left( 2\Lambda -\kappa ^{2}\lambda \right) \tau ^{2}\leq 8
\end{equation}%
should hold, which again is satisfied by any $\Lambda \leq \kappa
^{2}\lambda /2$, but would be clearly violated for $\Lambda >\kappa
^{2}\lambda /2$ at $\tau >\tau _{2}\equiv 2\sqrt{2/3\left( 2\Lambda -\kappa
^{2}\lambda \right) }$ .We summarize the results for $\rho $ in Table \ref%
{Table1}.

\begin{table}[h]
\caption{Domains of positivity, negativity and ill-definedness of $\protect%
\rho $ for various values of $\Lambda $. The constants are $\protect\tau %
_{1}\equiv 2/\protect\sqrt{3\Lambda }$ and $\protect\tau _{2}\equiv 2\protect%
\sqrt{2/3\left( 2\Lambda -\protect\kappa ^{2}\protect\lambda \right) }$.}
\label{Table1}%
\begin{equation*}
\begin{tabular}{|c||c|c|c|c|}
\hline
$\rho$ & $\tau <\tau _{1}$ & $\tau =\tau _{1}$ & $\tau _{1}<\tau \leq \tau
_{2}$ & $\tau >\tau _{2}$ \\ \hline\hline
$\Lambda \leq 0\;$ & $+$ & $+$ & $+$ & $+$ \\ \hline
$0<\Lambda \leq \frac{\kappa ^{2}\lambda }{2}$ & $+$ & $0$ & $-$ & $-$ \\ 
\hline
$\Lambda >\frac{\kappa ^{2}\lambda }{2}$ & $+$ & $0$ & $-$ & no real solution
\\ \hline
\end{tabular}%
\end{equation*}%
\end{table}
By inserting Eqs. (\ref{atau}) and (\ref{rhotau}) in (\ref{rhop2}) we obtain
the evolution of the pressure in cosmological time $\tau $%
\begin{equation}
\frac{p}{\lambda }=1-\frac{4+3\left( \kappa ^{2}\lambda -2\Lambda \right)
\tau ^{2}}{\left( 3\lambda \right) ^{1/2}\kappa \tau \sqrt{8+3\left( \kappa
^{2}\lambda -2\Lambda \right) \tau ^{2}}}\ ,  \label{ptau}
\end{equation}%
Obviously, neither $p$ is well-defined when $\rho $ is not.

\section{Cosmological evolution}

It is instructive to raise the question, what is the \textit{total effective
source} in the modified Einstein equation (\ref{modE}) for the analyzed
Swiss-cheese models. For this, we note that in terms of $\rho $, $p$, $a$,
the $4$-velocity $u^{a}$ and the metric $g_{ab}=-u_{a}u_{b}+a^{2}h_{ab}$,
the linear source term is%
\begin{equation}
T_{ab}=\rho u_{a}u_{b}+pa^{2}h_{ab}\ .
\end{equation}%
The non-linear source term (\ref{S}) can be expressed as \cite{Decomp} 
\begin{equation}
\widetilde{\kappa }^{4}S_{ab}=\kappa ^{2}\frac{\rho }{\lambda }\left[ \frac{%
\rho }{2}u_{a}u_{b}+\left( \frac{\rho }{2}+p\right) a^{2}h_{ab}\right] \ .
\label{fS}
\end{equation}%
By employing Eqs. (\ref{rhotau}) and (\ref{ptau}) the total effective source
of the modified Einstein equation becomes%
\begin{equation}
-\Lambda g_{ab}+\kappa ^{2}T_{ab}+\widetilde{\kappa }^{4}S_{ab}=\frac{4}{%
3\tau ^{2}}u_{a}u_{b}\ .  \label{effectivesource}
\end{equation}%
Therefore the FLRW regions effectively behave like filled with a dust with
energy density 
\begin{equation}
\rho ^{tot}=\frac{4}{3\kappa ^{2}\tau ^{2}}\ .  \label{dust}
\end{equation}%
\begin{figure}[tbp]
\includegraphics[height=5.5cm]{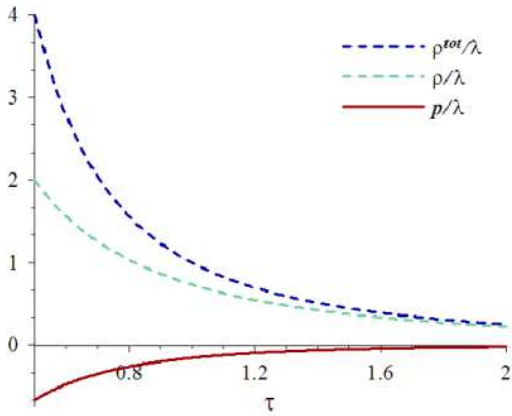}
\caption{(color online). Evolution in cosmological time (given in units $%
\left( 4/3\protect\kappa ^{2}\protect\lambda \right) ^{1/2}$) of $\protect%
\rho ^{tot}/\protect\lambda $ (upper curve), $\protect\rho /\protect\lambda $
(middle curve) and of $p/\protect\lambda $ (lower curve), for $\Lambda =0$.}
\label{Fig2}
\end{figure}
\begin{figure}[tbp]
\includegraphics[height=5.5cm]{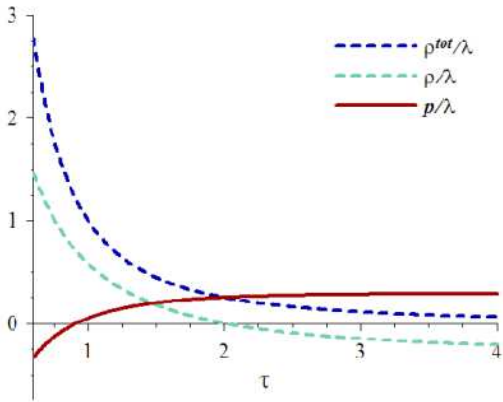}
\caption{(color online). As in Fig \protect\ref{Fig2} for $\Lambda =\protect%
\kappa ^{2}\protect\lambda /4$. We remark that the energy density turns
negative at $\protect\tau _{1}$. By contrast, the pressure becomes positive
during cosmological evolution. There is a time interval with both $\protect%
\rho $ and $p$ positive. }
\label{Fig3}
\end{figure}
\begin{figure}[tbp]
\includegraphics[height=6cm]{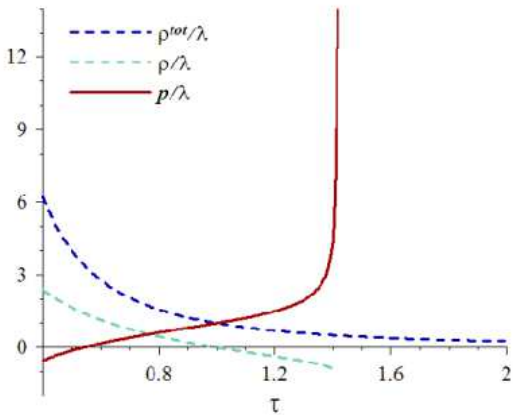}
\caption{(color online). As in Fig \protect\ref{Fig2}, for $\Lambda =\protect%
\kappa ^{2}\protect\lambda $. The evolution of $\protect\rho $ and $p$
follows the same pattern as in Fig \protect\ref{Fig2}, however at $\protect%
\tau =\protect\tau _{2}$ a pressure singularity appears. }
\label{Fig4}
\end{figure}

It is worth to stress that in the general relativistic Einstein-Straus model
the fluid is also dust. The comparison of the brane-world and general
relativistic Swiss-cheese models shows the same time dependence of the scale
factor, Eq. (\ref{atau}). Also, the time dependence of the dust energy
density in the Einstein-Straus model is identical with the time-dependence
of the effective total energy density (\ref{dust}) of the brane-world
Swiss-cheese model. In the brane-world Swiss-cheese model however the real
source is a perfect fluid, with energy density (\ref{rhotau}) and pressure (%
\ref{ptau}). The brane perfect fluid has three different possible behaviors,
depending on the value of the cosmological constant $\Lambda $ in the FLRW
regions, as shown in Table \ref{Table1}. For comparison we illustrate the
evolution of both of the energy densities (\ref{dust}) and (\ref{rhotau}),
together with the evolution of the pressure (\ref{ptau}) for these three
cases. Fig \ref{Fig2} is for $\Lambda =0$, Fig \ref{Fig3} for $\Lambda
=\kappa ^{2}\lambda /4$ and Fig \ref{Fig4} for $\Lambda =\kappa ^{2}\lambda $%
. These represent typical behaviors for cosmologies represented in the three
lines of Table \ref{Table1}.

In all cases the density of the cosmological fluid $\rho $ is less than the
total effective density $\rho ^{tot}$. The cosmological evolution (\ref{atau}%
), identical to that of the Einstein-Straus model, arises as consequence of
the \textit{modified dynamics} of the fluid in the brane-world scenario,
given by Eqs. (\ref{rhotau}) and (\ref{ptau}) and represented in Figs \ref%
{Fig2}-\ref{Fig4}.

How can one interpret the domains of negative energy density occurring
whenever $\Lambda >0$? In general relativity one can always redefine the
energy density and the pressure of an ideal fluid with the aid of a
cosmological constant such that the energy conditions are satisfied. Indeed,
a perfect fluid characterized by ($\rho ,~p$) together with a cosmological
constant $\Lambda $ are equivalent to an other perfect fluid with energy
density $\bar{\rho}=\rho +\Lambda /\kappa ^{2}$ and pressure $\bar{p}%
=p-\Lambda /\kappa ^{2}$ and no cosmological constant. Thus a negative
energy density $\rho $ can be turned positive by the above redefinition,
provided $\Lambda $ is a sufficiently large positive constant.

Due to the non-linear source terms $S_{ab}$ in the modified Einstein
equation (\ref{modE}) the general relativistic redefinition is not in order
any more. However in brane-world gravity an other transformation is possible
to the new fluid variables ($\bar{\rho},~\bar{p}$):%
\begin{eqnarray}
\frac{\bar{\rho}}{\lambda } &=&-1+\sqrt{1+\frac{2\Lambda }{\kappa
^{2}\lambda }+\frac{2\rho }{\lambda }\left( 1+\frac{\rho }{2\lambda }\right) 
}~,  \notag \\
\frac{\bar{p}}{\lambda } &=&1-\frac{1+\frac{2\Lambda }{\kappa ^{2}\lambda }+%
\frac{\rho }{\lambda }-\frac{p}{\lambda }\left( 1+\frac{\rho }{\lambda }%
\right) }{\sqrt{1+\frac{2\Lambda }{\kappa ^{2}\lambda }+\frac{2\rho }{%
\lambda }\left( 1+\frac{\rho }{2\lambda }\right) }}~,  \label{transf}
\end{eqnarray}%
with the result that the cosmological constant is absorbed into the new
fluid variables. Then the modified Friedmann and Raychaudhuri equations take
the form (\ref{Fried}) and (\ref{Raych}), but with $\Lambda =0$ and ($\bar{%
\rho},~\bar{p}$) in place of ($\rho ,~p$). Therefore the time evolution of
the new fluid variables is given by the $+$ branch of Eq. (\ref{rhotau}) and
Eq. (\ref{ptau}), both with $\Lambda =0.$ Consequently, the new energy
density stays positive during the whole evolution, while the new pressure is
negative, approaching zero at late times, as shown on Fig. \ref{Fig2}.

We conclude that in this brane-world model the evolution of the cosmological
fluid is realistic only for $\Lambda \leq 0$. For any positive $\Lambda $ a
negative energy density appears, which clearly represents exotic matter. In
these cases however the cosmological constant can be transformed out by the
redefinition (\ref{transf}) of the fluid variables. The transformation (\ref%
{transf}) at low energies ($\rho \ll \lambda $) approaches the general
relativistic transformation, while in the high energy / early universe
regime ($\rho \gg \lambda $) it approaches the identical transformation ($%
\bar{\rho},~\bar{p}$) $\approx $ ($\rho ,~p$), irrespective of the actual
value of $\Lambda $.

The question then comes, what is than the invariant meaning of $\Lambda $?
Originally it was introduced in the model as the cosmological constant in
the FLRW regions $\Lambda =\Lambda _{FLRW}$. If we also allow for a
cosmological constant $\Lambda _{void}$ in the voids (so that they become
Schwarzschild-de Sitter or Schwarzschild anti de Sitter regions instead
Schwarzschild black holes), then the junction conditions and the modified
gravitational dynamics on the brane give the model presented here, but with $%
\Lambda =\Lambda _{FLRW}-\Lambda _{void}$. Therefore if we allow for the
same cosmological constant in both regions on the brane, then $\Lambda =0$
and only the evolution presented on Fig \ref{Fig2} occurs. Otherwise we
encounter the pathological cases presented on Figs. \ref{Fig3} and \ref{Fig4}%
.

\section{Concluding Remarks}

We have explicitly constructed Swiss-cheese models by cutting out spheres of
constant comoving radius from the FLRW background with flat spatial sections
and filling them with Schwarzschild vacua. The latter are the brane sections
of black strings (cigars) from the bulk. The junction conditions and
cosmological evolution on the brane gave an expanding and decelerating
universe. The evolution in cosmological time of the energy density and
pressure of the cosmological fluid were given.

For $\Lambda \leq 0$ the energy density is positive and decreasing while the
pressure is always negative and increasing. The difference of the behavior
of the cosmological fluid in the brane-world Swiss-cheese model as compared
to the general relativistic Einstein-Straus model is robust in the early
universe. Later on, $\rho \rightarrow \rho ^{tot}$ and $p\rightarrow 0$.

For $0<\Lambda \leq \kappa ^{2}\lambda /2$ the energy density decreases from
positive values to negative values (at $\tau >\tau _{1}$). By contrast, the
pressure increases from negative values to positive ones, switching sign
when $\rho =\left( 2\lambda \Lambda \right) ^{1/2}/\kappa $. \textit{The
fluid evolves differently from the Einstein-Straus model} even in the long
run. We remark the existence of an epoch, when both the energy density and
the pressure of the cosmological fluid on the brane are positive. Even if we
exclude the regions with negative energy density from physical grounds, this
epoch can represent the static initial state of the static outcome of a
dynamic regime of the brane-world evolution .

For $\Lambda >\kappa ^{2}\lambda /2$ the evolution of the cosmological fluid
goes through the same sequence of epochs as for $0<\Lambda \leq \kappa
^{2}\lambda /2$, however at $\tau \rightarrow \tau _{2}$ the square root
from Eq. (\ref{rhotau}) tends to zero, thus $\rho \rightarrow -\lambda $ and 
$p\rightarrow \infty $. This resembles a sudden future singularity 
\cite{Dabrowski}, however it differs in an important aspect. In spite of
the infinite pressure at $\tau =\tau _{2}$, the acceleration given by the
Raychaudhuri equation (\ref{Raych}) remains finite, as can be seen from Eq. (%
\ref{adot2}). All other derivatives of the scale factor remain regular when
this \textit{pressure singularity }occurs. This is a novel type of
singularity arising in brane-wold scenarios and adds to the collection of
novel features encountered in the study of brane-world singularities. (For
example in Ref. \cite{ShtanovSahni} `quiescent'\ cosmological singularities
were presented, in which the matter density and Hubble parameter remain
finite, but all higher derivatives of the scale factor diverge as the
cosmological singularity is approached.)

In this paper we have studied the simplest brane-world Swiss-cheese model,
consisting of black strings (cigars) penetrating a cosmological FLRW brane
without dark radiation. Work in progress on more complicated models, arising
by the inclusion an asymmetry across the brane \cite{AsymSwissCheese} shows
that the generic features discussed here are shared by the whole class of
brane-world Swiss-cheese models.

Our analysis has shown that in the context of brame-world theories there is
not possible to construct a Swiss-cheese brane-world in which the
cosmological fluid is simple (like the dust in general relativity). Indeed,
even for the simplest case of $\Lambda =0$ (the same cosmological constant
in the FLRW regions and in the Schwarzschild voids), the fluid will obey the
condition for dark energy $\rho +3p<0$ at any $\tau \leq \left( 3\kappa
^{2}\lambda \right) ^{-1/2}$. In spite of the fluid behaving as dark energy,
the expansion is decelerated. This is due to the quadratic source term $%
S_{ab}$, which stays positive and dominates at high energies. Similar
features were also encountered in the study of the gravitational collapse on
the brane, when the exterior is static \cite{collapse}. The question comes,
whether a more reasonable behavior of the fluid will emerge if we relax the
condition $\mathcal{E}_{ab}=0$. This would mean to add tidal charge to the
brane black holes and to introduce black holes in the bulk as well. This
topic is currently under investigation \cite{tidalSwissCheese}.

\section{Acknowledgements}

This work was supported by OTKA grants no. T046939, TS044665, the J\'{a}nos
Bolyai Scholarship of the Hungarian Academy of Sciences and part of it done
during the Pomeranian Workshop in Fundamental Cosmology, Pobierowo, Poland
of the COSMOFUN\ collaboration. I thank the participants for a stimulating
atmosphere and the organizers for support. I also thank Roy Maartens for
comments on brane black holes.

\end{document}